\documentclass[preprint, showpacs,preprintnumbers,amsmath,amssymb,nofootinbib]{revtex4}

 \usepackage{epsf}
 \usepackage{graphicx}

\begin{document}
\newcommand{\be}[1]{\begin{equation}\label{#1}}
 \newcommand{\ee}{\end{equation}}
 \newcommand{\bea}{\begin{eqnarray}}
 \newcommand{\eea}{\end{eqnarray}}
 \def\disp{\displaystyle}

 \def\gsim{ \lower .75ex \hbox{$\sim$} \llap{\raise .27ex \hbox{$>$}} }
 \def\lsim{ \lower .75ex \hbox{$\sim$} \llap{\raise .27ex \hbox{$<$}} }

 \title{\Large \bf The growth factor of matter perturbations in an $f(R)$ gravity
}
   \author{Xiangyun Fu$^{1,2}$, Puxun Wu$^{1,2}$  and Hongwei Yu$^{1,2,}$$\footnote{e-mail:hwyu@hunnu.edu.cn}$}

\address{$^1$Department of Physics and Institute of  Physics, Hunan
Normal University, Changsha, Hunan 410081, China \\
$^2$Key Laboratory of Low Dimensional Quantum Structures and Quantum
Control of Ministry of Education, Hunan Normal University, Changsha,
Hunan 410081, China}

\begin{abstract}
The growth of matter perturbations in the $f(R)$ model proposed by
Starobinsky is studied in this paper. Three different parametric
forms of the growth index are considered respectively and
constraints on the model are obtained at both the $1\sigma$  and
$2\sigma$ confidence levels, by using the current observational data
for the growth factor. It is found, for all the three parametric
forms of the growth index examined, that the Starobinsky model is
consistent with the observations only at the $2\sigma$ confidence
level.

\end{abstract}


 \maketitle
 \renewcommand{\baselinestretch}{1.5}

\section{Introduction}\label{sec1}
The present cosmic accelerating expansion~\cite{Sne,Sne02,Sne03,Sne04,Sne05,Sne06,Sne07,Sne08,Sne09,Sne10, CMB,CMB02,CMB03,CMB04, SDSS,SDSS02,
ppeebles2,ppeebles202,ppeebles203,ppeebles204,ppeebles205,ppeebles206,ppeebles207,ppeebles208,ppeebles209,ppeebles210,ppeebles211,ppeebles212,ppeebles213} is one of the key challenges in  fundamental physics and
cosmology. There are basically two kinds of options to explain this
mysterious acceleration. One is the well known dark
energy~\cite{vsahni,vsahni02,vsahni03,vsahni04,vsahni05,vsahni06}, an energy component, which has a sufficient
negative pressure to induce a late-time accelerated expansion; the
other is the modified gravity, which originates from the idea that
our understanding of gravity is incorrect in the cosmic scale and
general relativity needs to be modified. One of the popular modified
gravities is the $f(R)$ theory~(see~\cite{ Nojiri20082,Nojiri2008202, Felice2010,Felice201002}
for a review), where $R$ is the Ricci scalar and $f(R)$ is an
arbitrary function of $R$. For an $f(R)$ model, its action takes the
form
 \be{action}
  S=\int {d^4}x\sqrt{-g}\bigg[{1\over16\pi G_N}f(R)+{L}_m\bigg]\,,
 \ee
where $g$ is the trace of the metric $g_{\mu\nu}$, $G_N$ is a bare
Newton gravity constant and $L_m$ is the Lagrangian of matter.
Considering a spatially flat Friedman-Lema\^{i}tre-Robertson-Walker
universe, whose metric is $ds^2=-dt^2+a^2(t)d{\bf x}^2$, and varying
the above action with respect to $g_{\mu\nu}$, one can obtain  \bea
3FH^{2} & = & 8\pi G_N ~(\rho_m +\rho_{rad})+\frac{1}{2}(FR-f)-3H\dot{F}~,\label{E1}\\
-2F\dot{H} & = & 8\pi G_N \left( \rho_m + \frac{4}{3}\rho_{rad}
\right)+ \ddot{F}-H\dot{F}~,\label{E2} \eea where $R=6(2 H^2+\dot
H)$,
 an over-dot stands for a derivative with respect to the cosmic time
$t$,  $H\equiv{\dot a\over a}$ is the Hubble parameter and $F\equiv
{df(R)\over dR}$.

Originally, Capozziello~\cite{capozziello} proposed an $f(R)$ model,
$f(R)=R-\alpha/R^m$ ($\alpha>0$, $m>0$), to explain the present
accelerating expansion. However,  this model was plagued with some
problems, which are related to the solar-system
constraints~\cite{chiba2}, the instabilities~\cite{dolgov}, a viable
cosmic evolution history with an accelerating
expansion~\cite{amendola1} and a standard matter-dominated
stage~\cite{carroll}. The main reason this model does not work is
that $f_{,RR}\equiv\partial^2f/\partial R^2<0$, which gives a
negative mass squared  for the scalaron field. Soon, the
aforementioned problems were solved, for example, the instabilities
and the inconsistence with the solar-system constraint were solved
in Refs.~\cite{Nojiri2003, Nojiri2004}, and the problem of matter
dominance was solved in Refs.~\cite{Capozziello2006, Nojiri2006, Nojiri2007}. Later, Amendola et al.~\cite{Amendola2007}
gave the conditions to obtain a viable $f(R)$ model. Some models
satisfying these conditions, the Starobinsky model, for an example,
have been proposed~\cite{Li2007,Li200702,Li200703,Li200704,Li200705,Li200706,Li200707, Amendola2007, starobinsky}.
Moreover, it is interesting to note that there are some
models~\cite{Nojiri2003, Nojiri20072, Nojiri20073, Nojiri2008,
Cognola2008} in $f(R)$ gravity, which can not only explain the
present accelerating expansion successfully, but, at the same time,
can also yield an inflation in the early era of our universe without
a scalar field.

Let us note that both the dark energy and $f(R)$ gravity can explain
the present accelerating expansion. However, although different
models can give the same late time expansion,  they may produce
different growths of matter perturbations~\cite{aastarobinsky}.
Thus,  the studies of the linear growth of matter
perturbations~\cite{dhuterer,dhuterer02,dhuterer03,dhuterer04,dhuterer05,dhuterer06,dhuterer07,dhuterer08,r12,r1202,r13,r14,r15,r16,r1602,r17,r18,r19,r22,r2202,
r28,r29,r30,r3002,r3003,r31,rgannouji1,rgannouji2,rgannouji202,rgannouji203,rgannouji204,rgannouji205,dpolarski,xiangyunfu,
xiangyunfu2,weihao,bboisseau} provide a particular method to
discriminate different models. Defining the growth function
$\delta(z)\equiv\delta\rho_m/\rho_m$ ($\rho_m $ is the energy
density of matter) and the growth factor $f\equiv{d\ln \delta\over
d\ln a}$, the authors in ~\cite{jnfry,jnfry02} found that $f$ can be
parameterized as
 \be{fommegam}
 f\simeq\Omega_m^\gamma,
 \ee
 where $\gamma$ is called the growth index and $\Omega_m$ is the
fractional energy density of matter. If $\gamma$ is treated as a
constant, its theoretical value can be obtained by expanding the
equation of $\gamma$ around $\Omega_m\simeq 1$, which is a good
approximation at the high redshift.  Then different models lead to
different theoretical values of
$\gamma$~\cite{r18,r19,r22,r2202,r28,r29,r30,r3002,r3003,r31,rgannouji2,rgannouji202,rgannouji203,rgannouji204,rgannouji205,rgannouji1,dpolarski,xiangyunfu,weihao,bboisseau},
for example, $\gamma_\infty\simeq6/11$~\cite{r18,r19} for
$\Lambda$CDM model and $\gamma_\infty \simeq11/16
$~\cite{r18,weihao} for flat DGP model. Therefore, it is possible to
distinguish them. By comparing the theoretical value of $\gamma$
with the observed one, one can hopefully single out the model which
is consistent with the observations.

However, the growth index is, in general, a function of redshift.
Some works have been done on the evolutionary form of $\gamma(z)$.
In Refs.~\cite{dpolarski, rgannouji1, rgannouji2,rgannouji202,rgannouji203,rgannouji204,rgannouji205, xiangyunfu}, the
authors studied $\gamma(z)$ with a linear expansion, $\gamma\approx
\gamma_0 + \gamma_0' z$, and found that this form gives a very good
approximation at the low redshift $z<0.5$ and for different models
$\gamma_0'$ is different.  Thus, an accurate measurement of
$\gamma_0'$ could provide another characteristic discriminative
signature to discriminate different models.  In
Refs~\cite{xiangyunfu2}, we proposed a parametrization
 $\gamma(z)=\gamma_0+\gamma_1 z/(1+z)$, and obtained that, for $w$CDM and
 DGP models, this form approximates the growth factor
 $f$ very well both at the low and high redshift regions.

In this paper, we aim to examine the growth factor of matter
perturbations in  $f(R)$ gravity and  we  take the Starobinsky
$f(R)$ model as an example. Let us note that the density
perturbations of the Starobinsky $f(R)$ model have been  studied
systematically in the literature~\cite{rgannouji2,rgannouji202,rgannouji203,rgannouji204,rgannouji205}. But what we plan
to do here is to examine different parametric forms of growth index
and study the observational constraints from the growth factor data.

\section{the Starobinsky's model}\label{sec12}

The Starobinsky's model has the form: \be{staro}
 f(R)= R+\lambda_s R_0\bigg[\bigg(1+{R^2\over
 R_0^2}\bigg)^{-n}-1\bigg],
 \ee
where $\lambda_s$ and $n>0$ are two positive constants, and $R_0$
corresponds essentially to the present value of  the Ricci scalar
$R$. This model has been studied in the
literature~\cite{rgannouji2,rgannouji202,rgannouji203,rgannouji204,rgannouji205,starobinsky} and it has been found that,
when $n\geq2$, all known the laboratory and Solar system tests of
gravity can be satisfied~\cite{starobinsky}. In this paper, we will
let $n=2$ for simplicity. Constant curvature solutions (for example:
de Sitter solution: $R=const=x_1 R_0>0$) are the roots of the
algebraic equation~\cite{starobinsky}
\begin{equation}
\label{desitter} Rf'(R)=2\,f(R).
\end{equation}
Substituting  the expression of $f(R)$ given in Eq.~(\ref{staro})
into the above equation, one can obtain
\begin{equation}\label{lambda2}
\lambda_s=\frac{x_1(1+x_1^2)^{n+1}}{2[(1+x_1^2)^{n+1}-1-(n+1)x_1^2]}.
\end{equation}
In order to satisfy the stability conditions of the system, the
following inequality must be satisfied~\cite{starobinsky}
\begin{equation}
\label{condition1} (1+x_1^2)^{n+2}>1+(n+2)x_1^2+(n+1)(2n+1)x_1^4.
\end{equation}
Setting $n=2$ and solving the above inequality, one gets
$x_1>\sqrt{\sqrt{13}-2}$,  which leads to $\lambda_s>0.94$.  We use
$\lambda_s=0.95$ in this paper, without loss of generality.

\section{ the growth of matter perturbations}
As shown in Refs.~\cite{Cognola2009, Nojiri2009}, the background
evolution of a viable $f(R)$ is very complicated. Here, we neglect
all higher derivative and non-linear terms, and we then obtain the
equation governing the growth of matter perturbations on subhorizon
scales as follows~\cite{bboisseau}
\begin{equation}
\label{denpert} \ddot{\delta}+2H\dot{\delta}-4\pi
G_{eff}\,\rho_m\delta=0,
\end{equation}
where $G_{eff}$ is an effective Newton gravity constant and for an
$f(R)$ model, it can be expressed as~\cite{tsujikawa}
\begin{equation}
\label{geff} G_{eff}={G_N\over F}\frac{1+4{k^2F'\over a^2
F}}{1+3{k^2F'\over a^2 F}}\,.
\end{equation}
Defining the growth factor $f\equiv d\ln\delta/d\ln a$, Eq.
(\ref{denpert}) becomes
\begin{equation}
\label{grwthfeq1} {d\; f\over d\ln
a}+f^2+{1\over2}\left(1-{d\ln\Omega_m\over d\ln
a}\right)f=\frac{3}{2}\frac{G_{eff}}{G_{N}}\Omega_m,
\end{equation}
Obviously, the growth factor is scale dependent, which leads to a
dispersion of growth index~\cite{Tsujikawa2009}. Here we consider
the wavenumber $k$ within the range
 \begin{eqnarray} 0.01~h~
Mpc^{-1}\lesssim k\lesssim 0.2~h~Mpc^{-1}\;,
\end{eqnarray} which is
relevant to the galaxy power spectrum~\cite{Percival2007}. In scale
smaller than $ 0.2~h~Mpc^{-1}$,  non-linear effects are obvious and
for scale larger than $0.01~h~ Mpc^{-1}$ the current observations
are not so accurate.

\subsection{a constant $\gamma$}
In this subsection, we discuss the parameterized form
$f\equiv\frac{d\ln\delta}{d\ln a}\simeq\Omega_m^\gamma$ with a
constant $\gamma$. Usually, the theoretical value of $\gamma$ can be
obtained by expanding the equation of $\gamma$ around
$\Omega_m\simeq 1$, which is a good approximation at the high
redshift. In principle, we can also obtain the theoretical value of
$\gamma$ by solving Eq.~(\ref{grwthfeq1}) numerically and using the
value of $\Omega_{m0}$ given by current observations. Since the
observational results on $\Omega_{m0}$ for Starobinsky's  model is
 not obtained yet, we use $\Omega_{m0}=0.278^{+0.024}_{-0.023}$ at
the $68\%$ confidence level given in Ref.~\cite{Wu2008}  with a
model independent method. Solving Eq.~(\ref{grwthfeq1}) to obtain
$f(0)$  numerically and using the relation
$f(0)=\Omega_{m0}^{\gamma_0}$ with $\Omega_{m0}$ taking the best fit
value $0.278$, we find $\gamma_0\simeq 0.42$, which seems to be
almost independent of the value of $k$.

\begin{figure}[htbp]
 \includegraphics[width=0.55\textwidth]{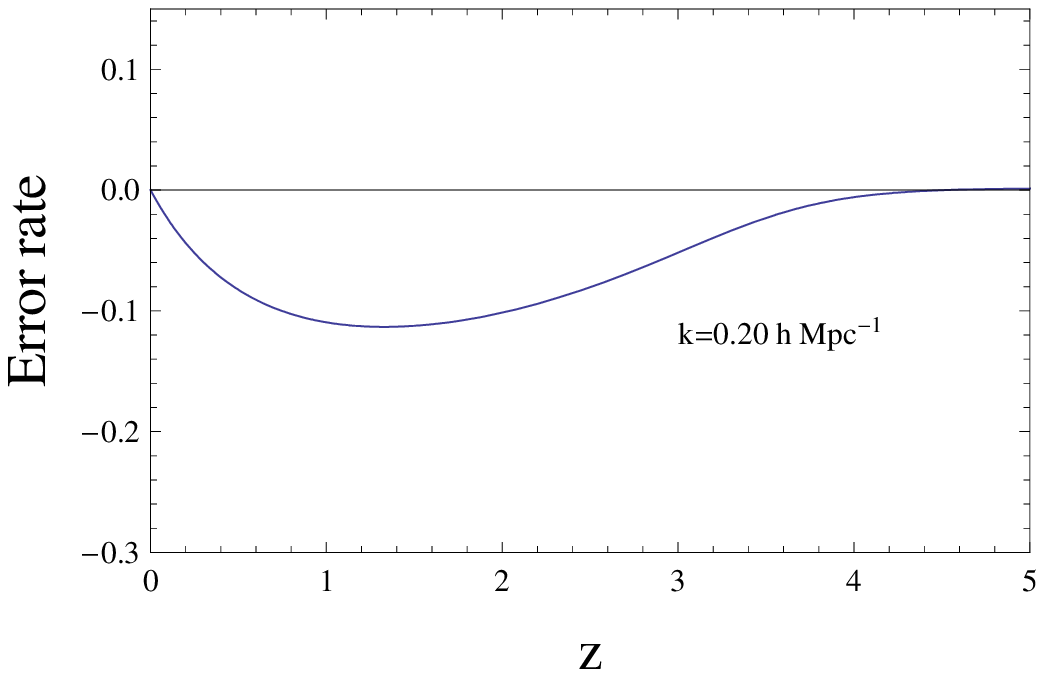}
\caption{\label{Fig1}
 The relative difference between the growth factor $f$ and
$\Omega_m^{\gamma_0} $with $\Omega_{m,0}=0.278$.  }
 \end{figure}

In order to discriminate different models with the growth factor, we
must compare the theoretical value and the observational one of
$\gamma$. The current observations give 12 data points of the growth
factor~\cite{guzzo,guzzo02, tegmark, ross, angela, mcdonald, viel2, viel1}.
Let us note that although the data given in Refs.~\cite{viel2,
viel1} are measured without `any' bias, other data points are
obtained by assuming a flat $\Lambda$CDM model with $\Omega_{m,0}$
taking a specific value, for example, $\Omega_{m,0}=0.25$ or $0.30$.
So, caution must be exercised when using these data.  With this
caveat in mind, it may still be worthwhile to apply the data to fit
 models~\cite{gongyungui, weihao, Zhang2008}. Using these 12 data,
we find that, for a constant $\gamma_0$ and $\Omega_{m0}=0.278$,
$\chi^2=4.6$ and $\gamma=0.63_{-0.14-0.33}^{+0.17+0.47}$ at the
$1\sigma$ and $2\sigma$ confidence levels. It is easy to see that
the Starobinsky's model is allowed only at the $2\sigma$ confidence
level. However, by comparing $f$ and $\Omega_m^{\gamma_0}$, we can
see that the error rate is larger than $10\%$ as shown in
Fig.~(\ref{Fig1}), which means that the result obtained with a
constant $\gamma$ may be biased. This bias arises from the fact that
$\gamma$ is a function of redshift instead of a constant. More
recently, the authors in Refs.~\cite{rgannouji2,rgannouji202,rgannouji203,rgannouji204,rgannouji205} discussed a
linearized form of $\gamma$ with $\gamma=\gamma_0+\gamma_1 z$, where
$\gamma_1\equiv\gamma_0'={d\gamma\over dz}(z=0)$. In the subsequent
subsection, we will examine this varying form of $\gamma$ in detail.

\subsection{$\gamma=\gamma_0 +\gamma_1 z$}

\begin{figure}[htbp]
 \includegraphics[width=0.55\textwidth]{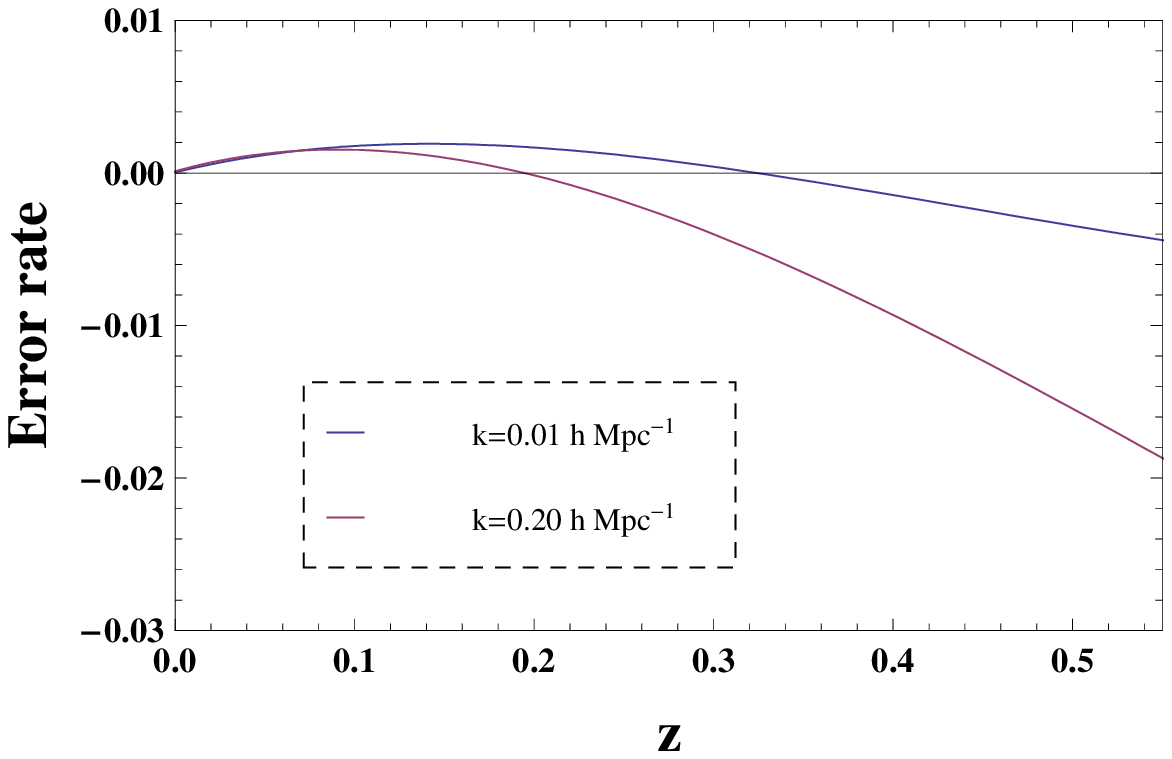}
\caption{\label{Fig2}
 The relative difference between the growth factor $f$ and
$\Omega_m^{\gamma_0+\gamma_1 z} $with $\Omega_{m,0}=0.278$.
}\end{figure}

This linearized form of $\gamma$ has been studied in the $w$CDM, DGP
and $f(R)$ gravity, and it gives a very good approximation at the
redshift region $z<0.5$. In Ref.~\cite{xiangyunfu}, we found that
the constraints on $\gamma_0$ and $\gamma_1$ from three low redshift
observational data cannot rule out the DGP model at $1\sigma$
confidence level. Here, we want to see what happens for  the
Starobinsky's model, where we have
\begin{equation}
 \label{gamma1}
 \gamma_1=[\ln\Omega_{m,0}^{-1}]^{-1}\bigg[-\Omega_{m,0}^{\gamma_0}-3(\gamma_0-{1\over2})(-1-{2\dot{H_0}\over3H_0^2})
 +{3\over2}{G_{eff}\over
 G_N}\Omega_{m,0}^{1-\gamma_0}-{1\over2}\bigg].
 \end{equation}
 When
$\Omega_{m,0}=0.278$, we obtain that $\gamma_0\simeq 0.41$. At the
same time, we find that, for different $k$, the variation of
$\gamma_1$ is small, for example, $\gamma_1$ varies from $-0.20$ to
$-0.24$ when $k$ is from $k=0.0.1~h~Map^{-1}$ to $0.2~h~Map^{-1}$.
In Fig.~(\ref{Fig2}), we give the relative difference between the
growth factor $f$ and $\Omega_m^{\gamma_0+\gamma_1 z}$ with
$\Omega_{m0}=0.278$ and find that, at low redshifts, the error is
below $2\%$, which means that this linearized form gives a better
approximation.

\begin{figure}[htbp]
\includegraphics[width=0.55\textwidth]{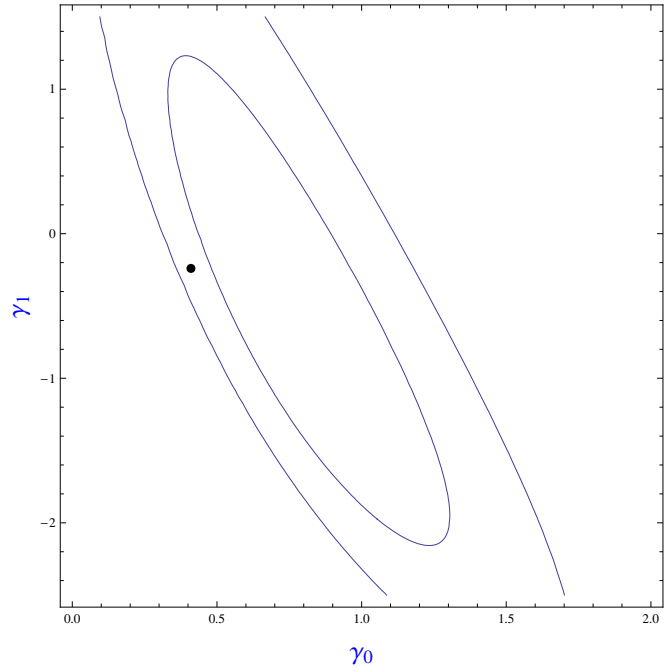}
\caption{\label{Fig3} The $1\sigma$ and $2\sigma$ contours of
$\gamma_0$ and $\gamma_1$ by fitting the  Starobinsky's model with
the three low redshift growth factor data.  }
 \end{figure}

Now we discuss the constraints on $\gamma_0$ and $\gamma_1$ from the
observations. Since this linearized form is valid in the low
redshifts,  only three low redshift data points can be used.
Fig.~(\ref{Fig3}) shows the results. From this figure, one can see
again that only at the  $2\sigma$ confidence level is the
Starobinsky's model  consistent with the observations and it can be
ruled out at the $1\sigma$ confidence level. This is in contrast
with the DGP model~\cite{xiangyunfu}.

However,  the approximate form  $f\simeq \Omega_m^{\gamma_0+
\gamma_1 z}$ is only valid at the low redshifts. In order to use all
the current observational data,  we need to find a new approximate
expression of $f$, which can give a good approximation in all
redshift regions,

\subsection{a new approximation of $f$}
From the Fig.~(2) in Ref.~\cite{rgannouji2,rgannouji202,rgannouji203,rgannouji204,rgannouji205}, which gives the
evolution of $f$, one can see that $f$ is larger than $1$ in the
region of $1<z<3.5$. Since in a flat universe $\Omega_m$ is always
less than one, according to the usual approximation $f\simeq
\Omega_m^{\gamma(z)}$ one cannot obtain $f>1$ if $\gamma(z)$ is
positive in the region $1<z<3.5$. Thus the usual parameterized form
of $f$ is hard to give a good approximation.

\begin{figure}[htbp]
\includegraphics[width=0.55\textwidth]{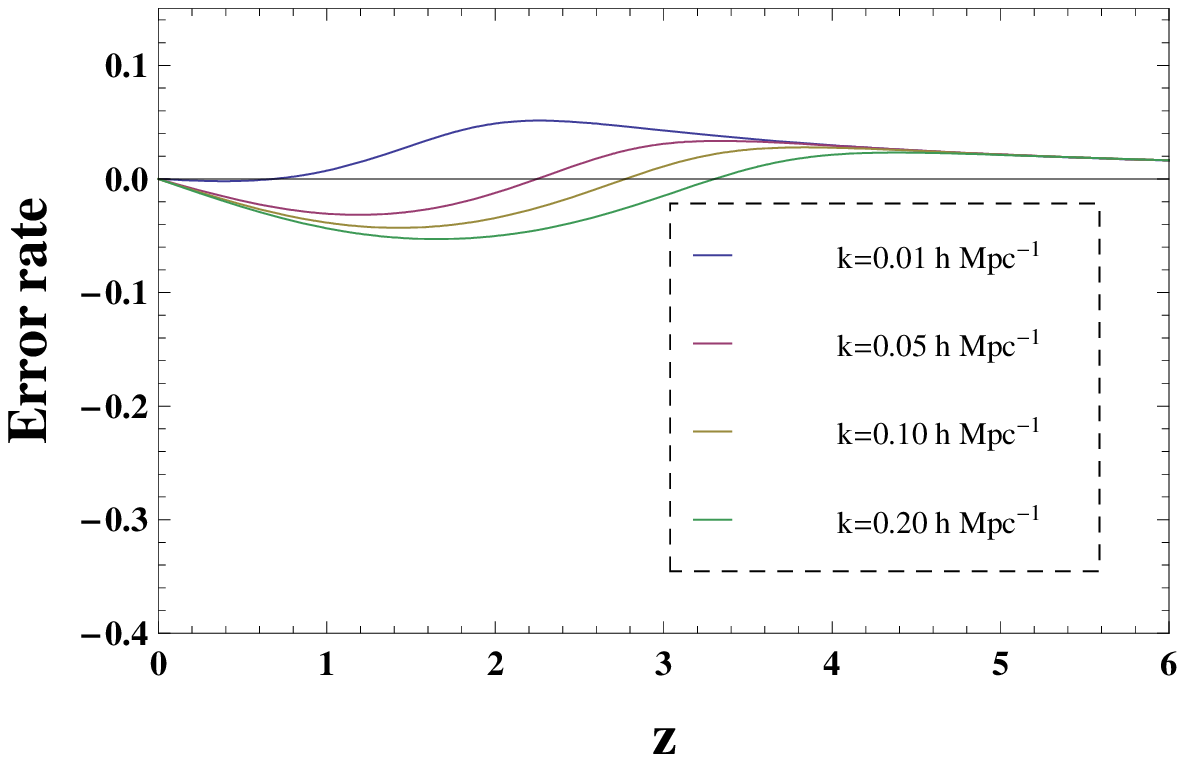}
\caption{\label{Fig4}The relative difference between the growth
factor $f$ and
$\big(1+\alpha\frac{1}{(1+z)^2+3}\big)\Omega_m^{\gamma_0}$ with
$\Omega_{m,0}=0.278$.  }
 \end{figure}

\begin{figure}[htbp]
\includegraphics[width=0.55\textwidth]{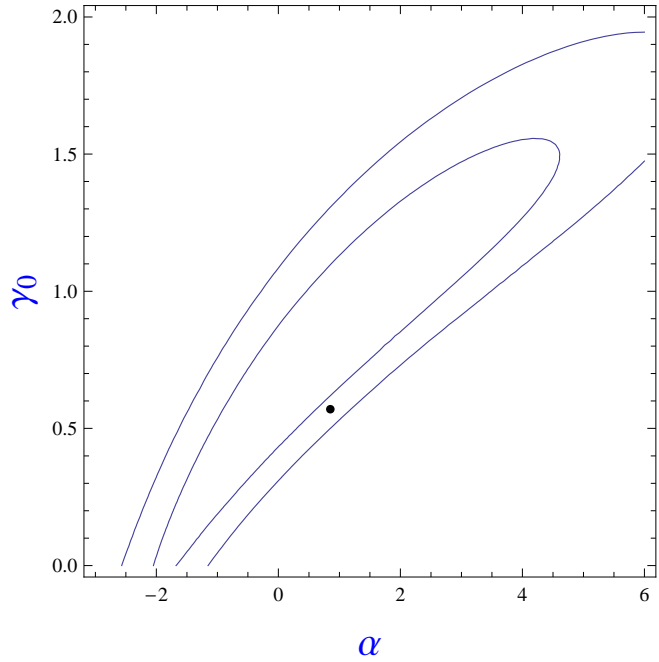}
\caption{\label{Fig5} The $1\sigma$ and $2\sigma$ contours of
$\gamma_0$ and $\alpha$ by fitting the  Starobinsky's model with the
current growth factor data.  }
 \end{figure}

From the definition of $G_{eff}$ and the fact that  $F$ is close to
one for $z>1$, one has at the redshift region $z>1$
\begin{eqnarray} \frac{G_{eff}}{G_N}\simeq 1+\frac{{k^2F'\over a^2 F}}{1+3{k^2F'\over a^2 F}}\;.
\end{eqnarray}   We assume an approximation of
$f$ by multiplying $\Omega_m^{\gamma_0}$ with a factor similar to
the above expression, i.e., we assume
\begin{eqnarray} f\simeq  \bigg(1+\alpha\frac{1}{(1+z)^2+3}\bigg) \Omega_m^{\gamma_0}\;,\end{eqnarray}
where $\alpha$ is a constant. We find that when $\alpha$ is about
equal to 0.85, for different wavenumbers $k$, the error rate, which
is defined as $ (1+\alpha\frac{1}{(1+z)^2+3} )\Omega_m^{\gamma_0}/f
-1$, is about less than $5\%$. The result is shown in
Fig.~(\ref{Fig4}). Therefore, with this new parametrized form of
$f$, one may use all the observational data.  After numerical
calculations, we obtain that $\gamma_0$ is about $0.57$, which seems
to be  almost independent of the value of wavenumbers k, when
$\alpha=0.85$. Using the 12 observational data points of the growth
factor, we place the constraints on $\alpha$ and $\gamma_0$, which
are shown in Fig.~(\ref{Fig5}). From this figure, we still find that
the Starobinsky's model is  allowed by the current observations only
at the $2\sigma$ confidence level.

\section{Conclusions}\label{sec4}
In this paper, we study the growth of matter perturbations in an
$f(R)$ model proposed by  Starobinsky. Firstly, we discuss the case
of a constant growth index. By comparing the theoretical value and
the observational one, we find that the Starobinsky model is allowed
by the current observations only at the $2\sigma$ confidence level.
However, in this case, the error rate between the growth factor $f$
and $\Omega_{m}^{\gamma_0}$ is larger than $10\%$, so, the result
obtained with a constant $\gamma$ may be biased. Then, a linear
expansion of growth index, $\gamma=\gamma_0+\gamma_1 z$, is studied,
which is valid at the low redshift region $z<0.5$ and gives a better
 approximation at these redshifts. With three low redshift
 observational
data, we find again that the Starobinsky model is allowed only at
the $2\sigma$ confidence level. Finally, in order to use all the
present data,  we propose a new approximate form of $f$, and show
that this new form gives a reasonable approximation both at low and
high redshift regions. For different scales, the largest error is
less than $5\%$. With this new proposed form of $f$, we still find
that the Starobinsky model is consistent with the observations only
at $2\sigma$ confidence level. So, our results seem to suggest that
although the Starobinsky $f(R)$ model is excluded by the current
growth factor data at  $1\sigma$ confidence level, it is still
allowed at $2\sigma$ level.

It should be pointed out that, in our discussion of the growth of
matter perturbations, the higher-derivative terms were discarded.
Recently, it has been found, that with the covariant perturbation
theory (see \cite{Carloni2009} for a recent review), which offers
the simplest way to describe the evolution of the perturbations,
these higher-derivatives terms can be kept in the analysis of matter
growth. So, it remains an interesting topic to examine what happens
when the effects of these terms are taken into account.

\section*{Acknowledgments}
Xiangyun Fu is grateful to R. Gannouji and Yungui Gong for their
very helpful communications. This work was supported in part by the
National Natural Science Foundation of China under Grants No.
10775050, 10705055 and 10935013, the SRFDP under Grant No.
20070542002, the Programme for the Key Discipline in Hunan Province,
 the FANEDD under Grant No. 200922, the Program for NCET (No.09-0144), and the
Hunan Provincial Innovation Foundation for Postgraduate No.
CX2009B101.

\end{document}